\documentclass[pss]{wiley2sp} 
\usepackage{amsmath}
\usepackage{units}

\begin{document}

\title{Nanosecond spin lifetimes in bottom-up fabricated bilayer graphene spin-valves with atomic layer deposited Al$_2$O$_3$ spin injection and detection barriers}

\titlerunning{Nanosecond spin lifetimes in bottom-up fabricated bilayer graphene spin-valves}

\author{%
  Marc Dr\"{o}geler\textsuperscript{\textsf{\bfseries 1}},
  Frank Volmer\textsuperscript{\textsf{\bfseries 1}},
  Maik Wolter\textsuperscript{\textsf{\bfseries 1}},
	Kenji Watanabe\textsuperscript{\textsf{\bfseries 2}},
	Takashi Taniguchi\textsuperscript{\textsf{\bfseries 2}},
	Daniel Neumaier\textsuperscript{\textsf{\bfseries 3}},
	Christoph Stampfer\textsuperscript{\textsf{\bfseries 1}},
	Bernd Beschoten\textsuperscript{\Ast,\textsf{\bfseries 1}}}

\authorrunning{Dr\"{o}geler et al.}

\mail{e-mail
  \textsf{bernd.beschoten@physik.rwth-aachen.de}}

\institute{%
  \textsuperscript{1}\,2nd Institute of Physics and JARA-FIT, RWTH Aachen University, D-52074 Aachen, Germany\\
  \textsuperscript{2}\,National Institute for Materials Science, 1-1 Namiki, Tsukuba, 305-0044, Japan\\
  \textsuperscript{3}\,Advanced Microelectronic Center Aachen (AMICA), AMO GmbH, D-52074 Aachen, Germany}

\received{XXXX, revised XXXX, accepted XXXX} 
\published{XXXX} 

\keywords{graphene, boron nitride, spin transport, Hanle precession.}

\abstract{%
\abstcol{%
We present spin transport studies on bi- and trilayer graphene non-local spin-valves which have been fabricated by a bottom-up fabrication method. By this technique, spin injection electrodes are first deposited onto Si$^{++}$/SiO$_2$ substrates with subsequent mechanical transfer of a graphene/hBN heterostructure. We showed previously that this technique allows for nanosecond spin lifetimes at room temperature combined with carrier mobilities which exceed $\unit[20,000]{cm^2/(Vs)}$. Despite strongly enhanced spin and charge transport properties, the MgO injection barriers in these devices exhibit conducting pinholes which still limit the measured spin lifetimes. We demonstrate that these pinholes can be partially dimini-
}{
ished by an oxygen treatment of a trilayer graphene device which is seen by a strong increase of the contact resistance area products of the Co/MgO electrodes. At the same time, the spin lifetime increases from 1~ns to 2~ns. We believe that the pinholes partially result from the directional growth in molecular beam epitaxy. For a second set of devices, we therefore used atomic layer deposition of Al$_2$O$_3$ which offers the possibility to isotropically deposit more homogeneous barriers. While the contacts of the as-fabricated bilayer graphene devices are non-conductive, we can partially break the oxide barriers by voltage pulses. Thereafter, the devices also exhibit nanosecond spin lifetimes.
 }}

\maketitle   

\section{Introduction}

Graphene has drawn a lot of attention in the past years thanks to its excellent optical, mechanical and electrical properties. Thanks to the small spin orbit coupling and hyperfine coupling it is also a very promising material in the field of spintronics. Spin lifetimes on the order of $\mu$s have been proposed \cite{Ertler2009}. However, first experiments using non-local graphene-based spin valves on Si/SiO$_2$ revealed typical spin lifetimes below $\unit[1]{ns}$ and electron mobilities below $\unit[10,000]{cm^2/(Vs)}$ \cite{Tombros2007,Han2009,Jozsa2009,Popinciuc2009,Han2010,Maassen2011,Avsar2011,Yang2011,PhysRevB.84.075453,PhysRevB.87.081405,abel:03D115,Kamalakar2014,1.4893578}. While the carrier mobilities could significantly be enhanced in suspended graphene devices or in devices in which the graphene sheet is in contact to hexagonal boron nitride (hBN), initial spin transport experiments on those devices still exhibited short spin lifetimes \cite{Guimaraes2012,Zomer2012}.

Later, it was demonstrated that the contacts are a bottleneck for the spin transport, e.g. it was shown that the measured spin lifetime scales with the contact area product ($R_\text{c}A$) over a wide range \cite{Volmer2013}, that an oxidation process of the graphene-electrode-interface can enhance the spin lifetime after the actual fabrication process \cite{Volmer2014} and that an increase in the electrode spacing (hence an increase in the ratio between graphene and electrode part) also yields longer spin lifetimes \cite{Kamalakar2015,2015arXiv150600472I}. It has also been shown that the injection barriers are often not homogeneous and exhibit pinholes \cite{2053-1583-2-2-024001}, which are likely a source of spin scattering \cite{Volmer2013}. But due to the inert surface of graphene it is difficult to grow continuous oxide layers without introducing artificial nucleation sites or modifying the graphene layer \cite{Wang2008,Dlubak2010}. In particular, this also holds for oxide barriers grown by atomic layer deposition (ALD). By this method, a precursor gas gets chemisorbed on the surface to form a self-terminated monolayer which is usually not obtained for inert surfaces like pristine graphene \cite{Puurunen2005,Yamaguchi2012}.

One way to diminish pinholes within the oxide barrier is subsequent oxidation of the whole device. The oxygen intercalates along the graphene-to-metal-oxide interface region and post-oxidizes the metal. With these oxygen treatments, spin lifetimes could be increased by a factor of 7, reaching $\unit[1]{ns}$ in the same device \cite{Volmer2014}. The tradeoff by this approach is a significant reduction of the carrier mobility.

To overcome these shortcomings, we recently introduced a bottom-up approach where the Co/MgO electrodes were not directly deposited onto the graphene layer but were rather fabricated onto Si$^{++}$/SiO$_2$ with the MgO as the surface layer. Subsequently, a graphene flake is picked-up by a hBN flake (van der Waals pick-up, similar technique to the one in Ref. \cite{Wang01112013}). Thereafter, the graphene/hBN stack is mechanically transferred on top of the predefined electrodes, i.e. onto the MgO surfaces of the electrodes. Using this method, carrier mobilities of $\unit[20,000]{cm^2/(Vs)}$ and spin lifetimes up to $\unit[3.7]{ns}$ at room temperature could be achieved \cite{Droegeler2014}. We note that even these devices seem to exhibit pinholes in the injection MgO barriers which can be probed indirectly by differential $V$/$I$-curves. Moreover, conductive scanning force microscopy measurements directly demonstrate the presence of conducting pinholes in these structures when growing Co/MgO electrodes by molecular beam epitaxy (MBE) \cite{tbp}. Despite the presence of pinholes, all devices exhibit nanosecond spin lifetimes. It is thus suggestive that even longer spin lifetimes could be achieved for homogeneous and pinhole-free injection barriers.

\begin{figure}[tb]
	\includegraphics{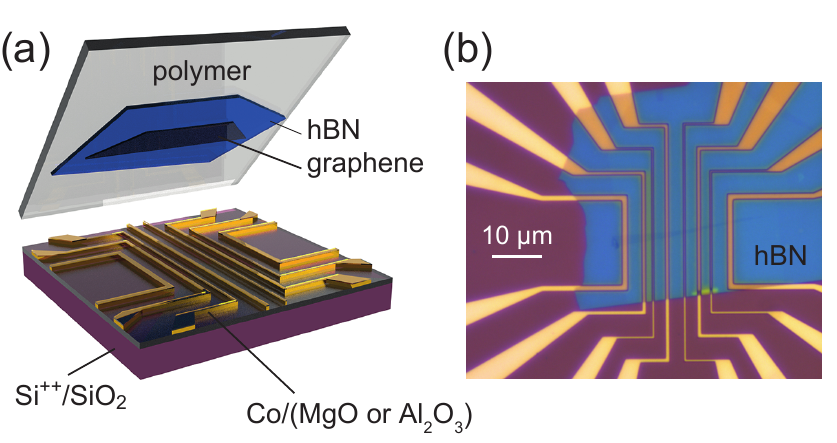}
	\caption{(Color online) (a) Schematic illustration of the device fabrication. The graphene flake is picked up from a plain Si/SiO$_2$ wafer by a stack of hBN on polymer. Thereafter, is placed on predefined Co/MgO or Co/Al$_2$O$_3$ electrodes. (b) Optical micrograph (top view image) of a complete device.}
	\label{fig:fig1}
\end{figure}

In this work, we investigate the contact and spin transport properties of graphene non-local spin valves using the bottom-up fabrication method described in Ref. \cite{Droegeler2014} for two sets of devices. In the first device, we use Co/MgO contacts grown by MBE and investigate the effect of oxidation on the contact and spin transport properties. We find that the post-oxidation of the MgO barrier yield both a larger contact resistance area product and a doubling of the spin lifetime from $\unit[1]{ns}$ to $\unit[2]{ns}$ indicating that even in devices with nanosecond spin lifetime the injection/detection barriers are not ideal and still limit the measured spin lifetimes. In a second device we use an Al$_2$O$_3$ barrier grown by ALD on top of MBE grown Co electrodes in order to achieve more homogeneous barriers. Here, ALD growth is expected to be more valuable for graphene spin transport devices as the ALD layer is not directly grown on graphene (which only can be achieved by introducing an additional adhesion layer \cite{Yamaguchi2012}) but rather on the ferromagnetic Co layer. Using ALD grown Al$_2$O$_3$ barriers we achieve similar nanosecond spin lifetimes and carrier mobilities in bilayer graphene as for devices with MgO barriers, despite the fact that pinholes were generated in the oxide barriers during the electrical contacting of the device.

\section{Device fabrication}

The device fabrication follows the method as described in Ref. \cite{Droegeler2014}. It consists of two main steps. In a first step, we prepare Co/MgO or Co/Al$_2$O$_3$ electrodes onto a Si$^{++}$/SiO$_2$ substrate, where the Si$^{++}$ can be used as a back gate. The electrodes are defined by standard electron beam lithography and metalized by MBE. We use a $\unit[40]{nm}$ thick Co layer for all devices. Afterward, we deposit $\unit[1]{nm}$ of MgO on top of Co in the same MBE chamber. Alternatively, we use Al$_2$O$_3$ as an injection and detection barrier. For these devices, we performed a lift-off process after Co deposition and transferred the sample to an ALD chamber where we grew a $\unit[0.88]{nm}$ thick Al$_2$O$_3$ layer using a trimethyl aluminum precursor and water vapor to oxidize the precursor. We note that the Al$_2$O$_3$ layer completely covers the Co electrodes (from top and from the sides) and the SiO$_2$ surface while for the previous process the MgO layer is only deposited on top of the Co layer. In a second step, we use exfoliated hBN on a polymer membrane to pick-up graphene which we previously exfoliated onto a plain Si/SiO$_2$ chip for both devices. Subsequently, the graphene/hBN stack is deposited onto both types of predefined electrodes as illustrated in Fig. \ref{fig:fig1}a. Thereafter, the polymer membrane is dissolved in acetone. An optical micrograph of a finished device is shown in Fig. \ref{fig:fig1}b. All subsequent spin and charge transport measurements were performed at room temperature under vacuum condition.

\begin{figure}[tb]
	\includegraphics{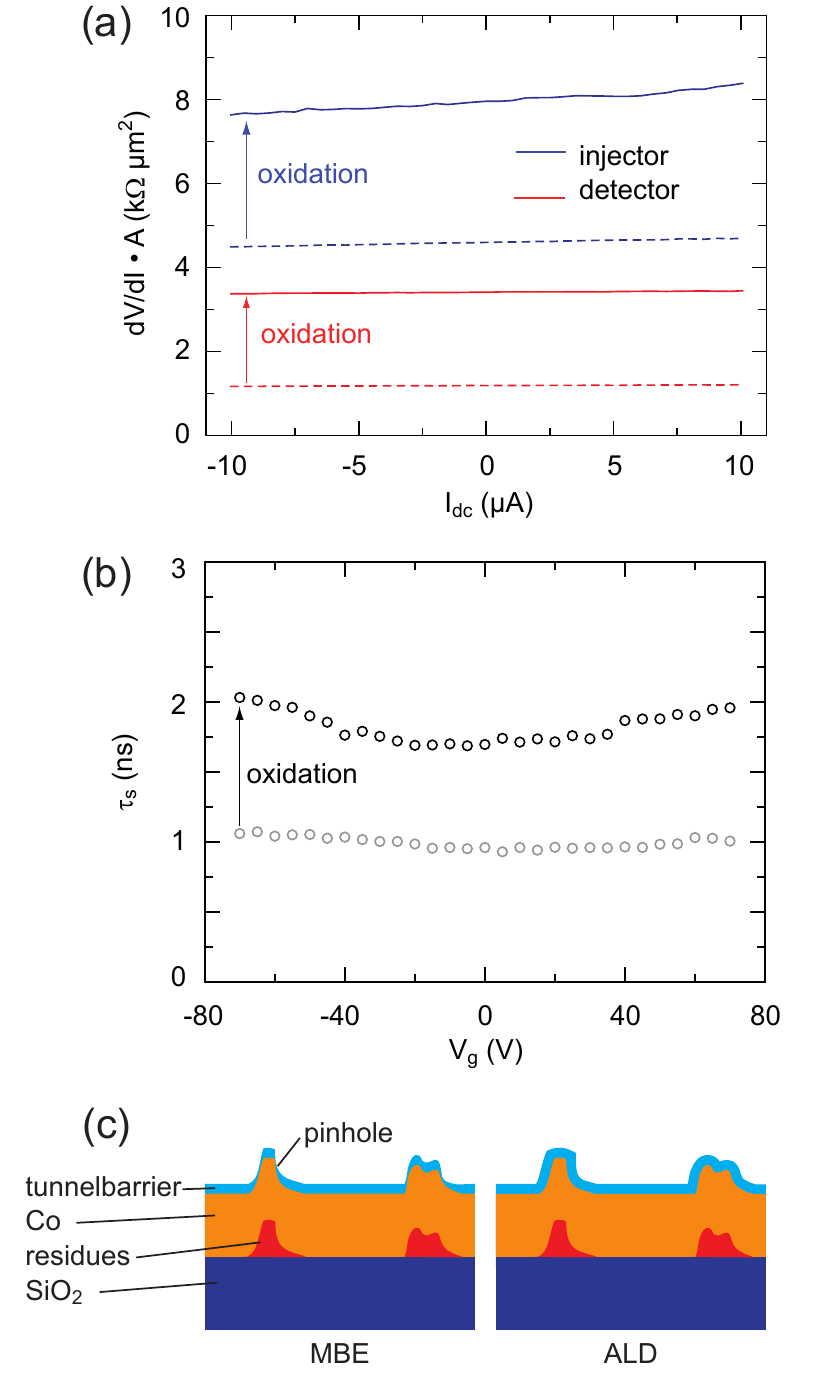}
	\caption{(Color online) (a) Differential $V$/$I\times A$ curves of spin injection and detection contacts before (dashed lines) and after oxidation (solid lines). (b) Corresponding back gate dependent spin lifetimes as extracted from non-local Hanle measurements. (c) Illustration of possible pinhole forming due to residues on the substrate and non-uniform growth of the MgO barrier and its possible circumvention by ALD growth.}
	\label{fig:fig2}
\end{figure}

\section{Oxidation of MgO barrier in trilayer graphene spin transport device}

We first discuss the influence of oxygen treatment on both the contact resistance area products ($R_\text{c}A$) and the spin lifetimes ($\tau_\text{s}$) for Co electrodes covered with a $\unit[1]{nm}$ thick MgO layer as grown by MBE. For this experiment we picked-up a trilayer graphene flake which was confirmed by Raman spectroscopy (data not shown).


At first, the contacts are characterized by measuring differential $V$/$I$ curves of the spin injection and detection electrodes. For comparison, these curves are normalized to their respective contact areas $A$. As all electrodes can be contacted from both sides of the graphene flake (see Fig. \ref{fig:fig1}b), we use a four-terminal measurement of the contact resistances as described in Ref. \cite{Volmer2013} where the measured voltage drops across the MgO barrier. In these measurements, we use a dc current which is modulated by a small ac current and detect the ac filtered voltage drop. A good tunneling contact would yield a non-linear $I-V$ curve which results in a distinct cusp in the d$V$/d$I$ characteristic. In contrast, the measured d$V$/d$I \times A$ curves of both contacts as a function of $I_{dc}$ remain completely flat indicating transparent and not tunneling contacts (Fig. \ref{fig:fig2}a dashed lines).

The value of $\unit[4.6]{k\Omega\cdot\mu m^2}$ for the injector and $\unit[1.3]{k\Omega\cdot\mu m^2}$ for the detector is slightly less than typical values in our previous study \cite{Droegeler2014}. We note that devices which were fabricated by the bottom-up approach exhibit transparent contacts for $R_\text{c}A$ products less than   $\unit[20]{k\Omega\cdot\mu m^2}$. After storing the device under ambient condition for 60 days we observe a strong increase of the $R_\text{c}A$ products by 180\% for the detector and by 70\% for the injector (Fig. \ref{fig:fig2}a solid lines). The respective d$V$/d$I \times A$ curves are still more or less flat demonstrating that the oxygen treatment cannot turn the contact properties from transparent to tunneling behavior in these devices. In gate dependent measurements of the graphene resistance we observe a strong reduction of the electron mobility from $\unit[8,300]{cm^2/(Vs)}$ before oxidation to $\unit[2,700]{cm^2/(Vs)}$ after oxidation. This overall trend, i.e. a decrease of carrier mobility with increasing $R_\text{c}A$ values by barrier oxidation, is in agreement to what has been measured for graphene on SiO$_2$ with top electrodes.\cite{Volmer2013}

\begin{figure*}[tb]
\centering	
\includegraphics{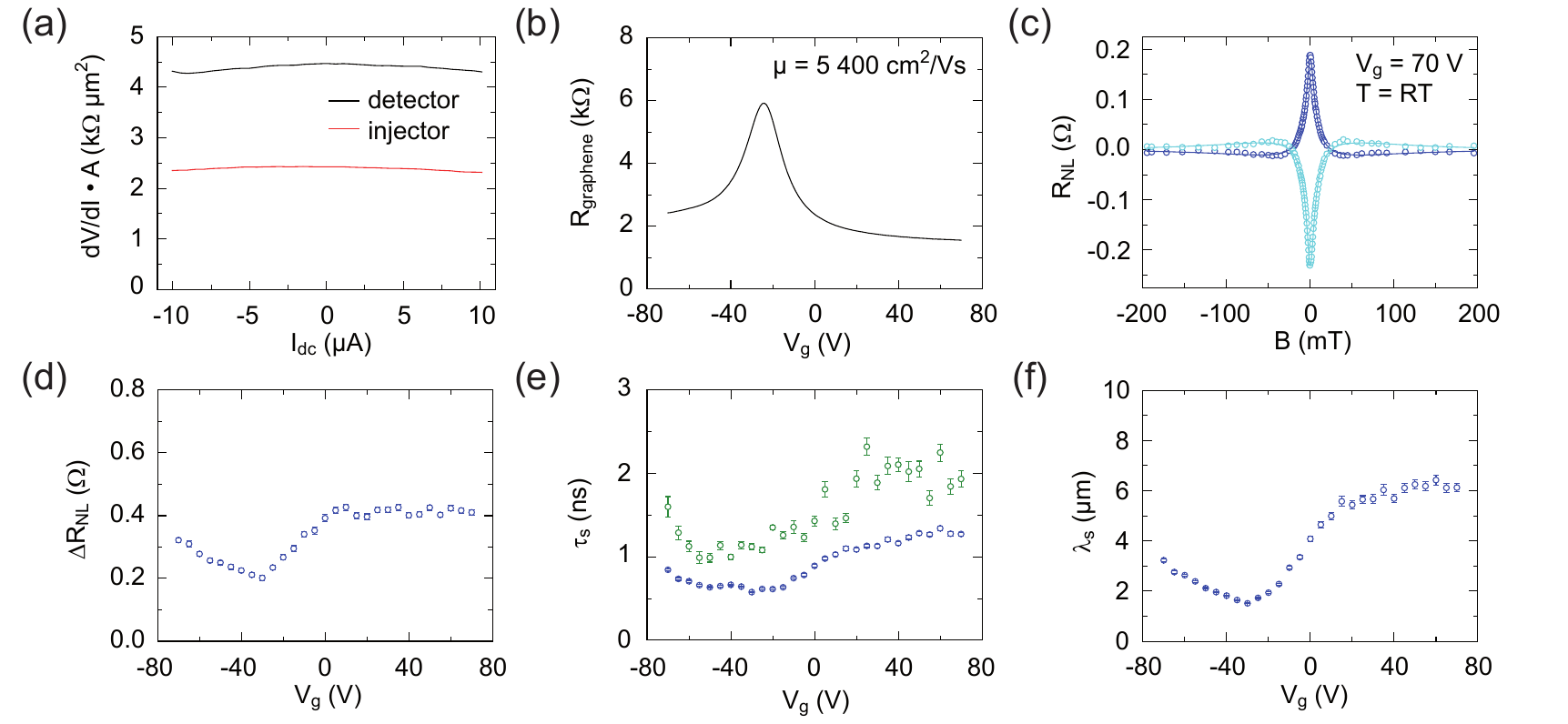}
	\caption{(Color online) Data set acquired from a bilayer samples with an ALD grown Al$_2$O$_3$ tunnel barrier. (a) Differential contact resistance normalized by the contact area of the injector and detector contact. (b) Dirac curve measured in four terminal geometry of the spin transport region. (c) Hanle depolarization curve for parallel and antiparallel alignment of the injector and detector at $V_\text g = \unit[70]{V}$. Gate dependence of (d) the spin signal, (e) the spin lifetime and (f) the spin diffusion length. Additionally, in (e) the spin lifetime of a different region on the same device is plotted (green symbols) which showed overall longer spin lifetimes.}
	\label{fig:fig3}
\end{figure*}

To further explore the effect of oxidation on the spin transport properties we extract spin lifetimes $\tau_s$ from non-local Hanle depolarization measurements of the spin resistance in perpendicular magnetic fields (see also Fig. \ref{fig:fig3}c).~\cite{Jedema2002,Tombros2007,Lou2007} We record Hanle curves for both parallel and antiparallel alignments of neighboring Co injector and detector electrodes for different backgate voltages. For the as-fabricated device, the spin lifetime exceed $\unit[1]{ns}$ and shows almost no backgate dependence. This has previously also been observed in bottom-up fabricated bi- and trilayer graphene spin-valves \cite{Droegeler2014}. After oxidation, the spin lifetime has increased to $\unit[1.7]{ns}$ at $V_\text g=\unit[0]{V}$. Interestingly, the spin transport regains gate tunability with spin lifetimes reaching $\unit[2]{ns}$ for both electron ($V_\text g>0$) and hole doping ($V_\text g<0$). Similar to our previous study \cite{Volmer2014}, we attribute the strong increase in spin lifetime to the improvement of the barrier quality which is seen by the increase of the $R_\text cA$ products during oxygen treatment. This finding suggests that even for bottom-up devices with spin lifetimes in the range of $\unit[1-2]{ns}$ the spin transport parameters such as the spin lifetime is limited by the contact properties.

Since oxidation of the contacts has a large impact on the measured spin lifetimes and a flat d$V$/d$I$ characteristic of the contacts is observed even after oxidation, we conclude that the spin injection and detection process is still governed by conducting pinholes. These pinholes can occur either due to an island formation of the MgO during growth or residual particles on the substrate which lead to an uneven Co surface \cite{Volmer2015}. If the resulting corrugation is larger than the MgO layer thickness the side of the corrugation may not be completely covered by the directional MBE growth of MgO which favors metallic contact to graphene, i.e. a pinhole. This case is illustrated in the left panel of Fig. \ref{fig:fig2}c. We therefore test our bottom-up approach with ALD growth of Al$_2$O$_3$ barriers.

\section{Fabrication of Al$_2$O$_3$ barriers by atomic layer deposition in bilayer graphene spin transport device}

ALD grown Al$_2$O$_3$ barriers may suppress pinhole formation to a large extend as the ALD process yields a complete coverage of the whole surface layer including the sidewalls of possible corrugations. Therefore, we expect more homogeneous injection barriers (see Fig. \ref{fig:fig2}c right panel) and hence even longer spin lifetimes. For this purpose, we used a bilayer graphene (BLG) flake and a $\unit[0.88]{nm}$ thick Al$_2$O$_3$ barrier.

The contact resistances for the as-fabricated device were in the order of several $\unit[]{G\Omega}$ which did not allow any spin or charge transport measurements. We therefore applied a higher voltage to enforce current flow through the device by breaking down the oxide barriers. We attribute this initial high contact resistances to an additional oxidation of the Co layer which naturally forms after removing the sample from the MBE chamber, performing lift-off and transferring it into the ALD chamber. Together with the Al$_2$O$_3$ layer, this cobalt oxide layer leads to an effectively thicker tunnel barrier. After applying the voltage pulse train, the barriers broke down most likely by creating artificial conducting pinholes. This can be seen from the respective d$V$/d$I$ characteristics in Fig. \ref{fig:fig3}a. As explained above, the almost flat curves again indicate transparent injection and detection barriers with comparable d$V$/d$I \times A$ values as for the device presented in the last section.

First, we determine the backgate characteristic of the graphene resistance which is shown in Fig. \ref{fig:fig3}b. The shift of the charge neutrality point (CNP) towards negative gate voltages results from electron doping of the graphene. Since the graphene layer bends down to the Al$_2$O$_3$ covered substrate in between the respective injection and detection electrodes, the observed doping is likely to be substrate-induced as already seen in Ref. \cite{Droegeler2014}. Either the doping is due to the Al$_2$O$_3$ itself or defects in the underlying SiO$_2$ layer interact with the graphene over the very thin ($\unit[0.88]{nm}$) Al$_2$O$_3$ layer.

The mobility $\mu$ is calculated from the slope of the gate dependent conductance $\sigma=1/\rho$ using $\partial \sigma/\partial n = \mu \times \text e$. We extract a room temperature electron mobility of $\unit[5,400]{cm^2/(Vs)}$. This value is by a factor of approximately 4 lower than what has been achieved in similar devices with MgO barriers. However, a different region of the same device has a mobility of $\unit[24,000]{cm^2/(Vs)}$. So far it is not clear whether this variation in mobility is caused by fabrication-induced impurities or is somehow linked to the Al$_2$O$_3$ barrier.

\begin{figure}[tb]
	\includegraphics{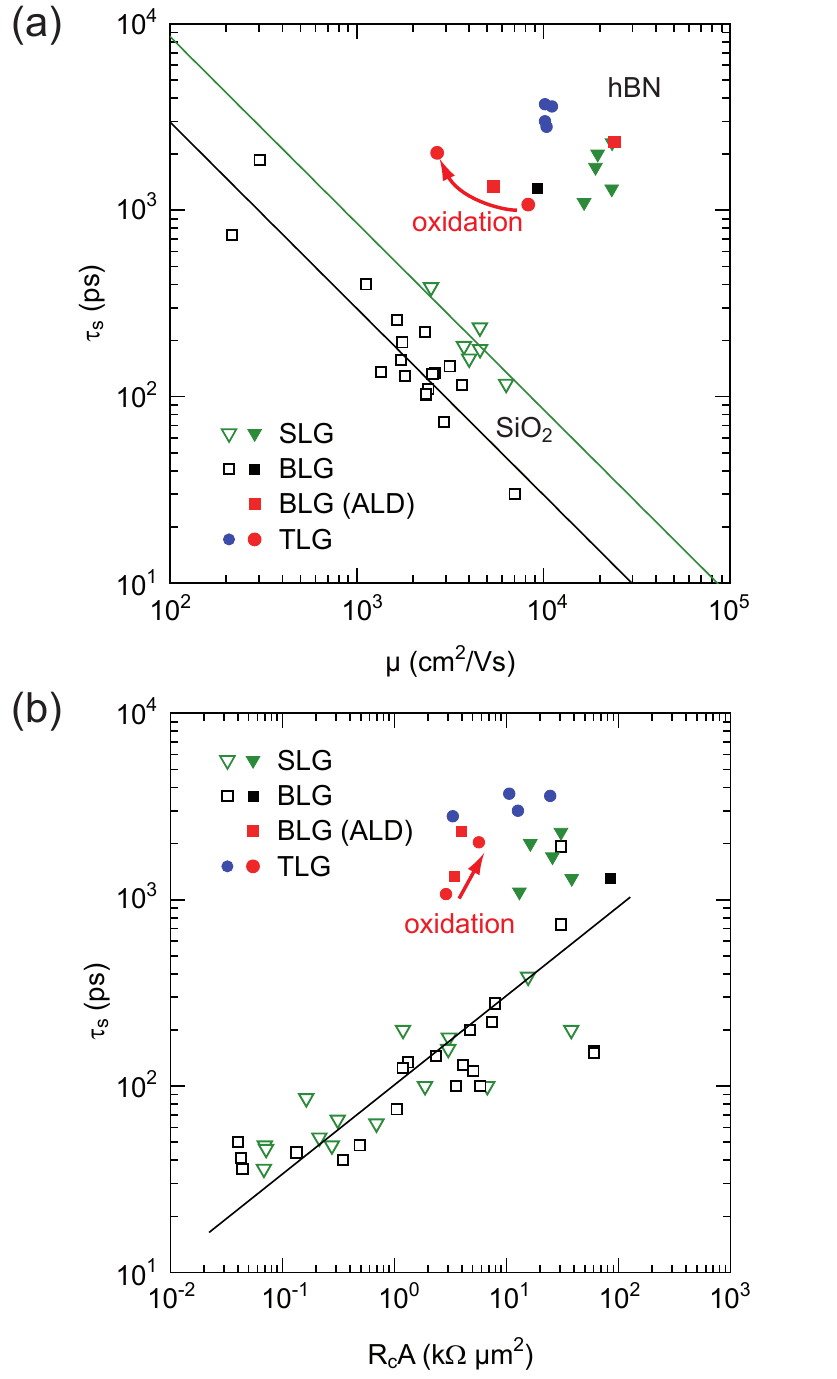}
	\caption{(Color online) (a) Room temperature spin lifetime versus electron mobility for non-local spin valve devices. The filled red symbols represent the data discussed in this work whereas the green, black and blue filled symbols are previous devices measured for the same bottom-up fabrication technique (taken from Ref. \cite{Droegeler2014}). The open symbols represent data obtained for graphene on SiO$_2$ (taken from Refs. \cite{Yang2011} and \cite{Volmer2013}). The solid lines illustrate the $1/\mu$ dependence for SLG and BLG observed previously on devices which were fabricated on SiO$_2$ by the conventional top-down approach.  (b) Spin lifetime versus contact-resistance-area products of respective injection and detection electrodes at room temperature. For comparison, we included all previous results from Ref. \cite{Volmer2013} on SLG and BLG devices that were fabricated by the conventional top-down method. The solid line is a guide to the eye.}
	\label{fig:fig4}
\end{figure}

In the following, we focus on the spin transport properties. The measured non-local Hanle signal is shown in Fig. \ref{fig:fig3}c for parallel and antiparallel alignments of the respective injector and detector electrodes at $V_\text{g} = +\unit[70]{V}$. We subtracted a parabolic background signal which results from charge accumulations in the non-local voltage \cite{2053-1583-2-2-024001}. The non-local spin signal $\Delta R_\text{NL}$ is given by the resistance difference at $B = \unit[0]{T}$. Fig. \ref{fig:fig3}d depicts the backgate dependence of $\Delta R_\text{NL}$. It shows a minimum at the CNP and increases for both electron and hole doping. For large positive gate voltages (large electron densities), the spin signal becomes constant. According to Han et al. \cite{Han2010} this dependence indicates transparent or semi-transparent contacts which is in agreement with the d$V$/d$I\times A$ characteristic and the breakdown of the barriers.

Despite the voltage treatments of the contact, we observe spin lifetimes of $\unit[1.3]{ns}$ at large positive gate voltages (see blue data points in Fig. \ref{fig:fig3}e). Surprisingly, this value is even slightly larger than what has seen before for BLG fabricated using the bottom-up approach. In the region with an electron mobility of $\unit[24,000]{cm^2/(Vs)}$ we even achieve a spin lifetime of more than $\unit[2]{ns}$ at large gate voltages (Fig. \ref{fig:fig3}e green symbols). The overall gate dependence is similar in both regions. The extracted spin diffusion length $\lambda_\text{s}$ is plotted in Fig. \ref{fig:fig3}f for the region with the lower mobility. Here values of $\unit[6.4]{\mu m}$ for high positive gate voltages are comparable to values achieved with MgO tunnel barriers.

\section{Conclusion}

Our results are summarized in Fig. \ref{fig:fig4} where we show both data from the present study (red symbols) together with data previously obtained on both the bottom-up devices (green, black and blue filled symbols) and the top-down devices (green and black open symbols). Fig. \ref{fig:fig4}a shows the respective spin lifetimes as a function of electron mobility on a log-log scale. All devices of the present study show comparable spin lifetimes and mobilities in comparison to our previous bottom-up devices.
The fact that the ALD devices exhibit similar transport characteristics after breaking-down its tunnel barriers
indicates that the spin properties are currently mainly governed by the conducting pinholes within the barriers and not by the electronic properties of the bare oxide material.

In Fig. \ref{fig:fig4}b we plot the spin lifetimes versus the R$_\text c$A value on a log-log plot. In comparison to devices which were fabricated by the top-down approach, all bottom-up devices exhibit overall larger R$_\text c$A values and, at the same time spin lifetimes above $\unit[1]{ns}$. For the latter devices there is no clear dependence of $\tau_s$ on the R$_\text c$A values.
Nevertheless, the increase of both $\tau_s$ and R$_\text c$A in the same device after oxidation clearly indicates that even in these devices the contacts have significant influence on the spin lifetime.

Finally, we want to stress that the ALD devices with Al$_2$O$_3$ barriers exhibit very similar device performances as all other bottom-up devices with MBE-grown MgO barriers after the breakdown of their barrier by voltage pulses. With the goal to fabricate homogeneous tunneling barriers the ALD process is thus advantageous over MBE growth under the premise that the native oxide (CoO in our case) of the underlying metal layer is removed prior to Al$_2$O$_3$ deposition.

\begin{acknowledgement}
The research leading to these results has received funding from the European Union Seventh Framework Programme under grant agreement n${^\circ}$604391 Graphene Flagship.
\end{acknowledgement}

%
%

\bibliographystyle{pss}
\bibliography{bibliography}

\end{document}